\batchmode
\makeatletter
\def\input@path{{"/home/jacob/Documents/Work/My Papers/2023-Stochastic Processes and Quantum Theory/"}}
\makeatother
\documentclass[11pt,twoside,english]{article}
\usepackage[T1]{fontenc}
\usepackage{textcomp}
\usepackage[latin9]{inputenc}
\usepackage{color}
\definecolor{lyxboxbgcolor}{rgb}{0.980469, 0.941406, 0.902344}
\usepackage{babel}
\usepackage{amsmath}
\usepackage{amssymb}
\usepackage{geometry}
\usepackage{setspace}
\usepackage[pdftex,pdfusetitle,
 bookmarks=false,
 breaklinks=false,pdfborder={0 0 0},pdfborderstyle={},backref=false,colorlinks=false]
 {hyperref}

\makeatletter

\usepackage{endnotes}

\let\originalleft\left
\let\originalright\right
\renewcommand{\left}{\mathopen{}\mathclose\bgroup\originalleft}
\renewcommand{\right}{\aftergroup\egroup\originalright}

\def\smalloverbrace#1{\mathop{\vbox{\m@th\ialign{##\crcr%
      \noalign{\kern3\p@}%
      \tiny\downbracefill\crcr\noalign{\kern3\p@\nointerlineskip}%
      $\hfil\displaystyle{#1}\hfil$\crcr}}}\limits}

\def\smallunderbrace#1{\mathop{\vtop{\m@th\ialign{##\crcr
   $\hfil\displaystyle{#1}\hfil$\crcr
   \noalign{\kern3\p@\nointerlineskip}%
   \tiny\upbracefill\crcr\noalign{\kern3\p@}}}}\limits}

\DeclareMathAlphabet{\mymathbb}{U}{bbold}{m}{n}



\makeatother

\begin{document}
\title{A Deflationary Account of Quantum Theory and its Implications for
the Complex Numbers}
\author{Jacob A. Barandes\thanks{Departments of Philosophy and Physics, Harvard University, Cambridge, MA 02138; jacob\_barandes@harvard.edu; ORCID: 0000-0002-3740-4418}
\thanks{This is the accepted manuscript, published in \emph{Philosophy of Physics} (DOI: 10.1017/psa.2026.10245)}}
\date{}

\maketitle

\begin{abstract}
Why does quantum theory need the complex numbers? With a view toward
answering this question, I argue that the usual Hilbert-space formalism
is a special case of the general method of Markovian embeddings. I
then describe the \textquoteleft indivisible interpretation\textquoteright{}
of quantum theory, according to which a quantum system can be regarded
as an \textquoteleft indivisible\textquoteright{} stochastic process
unfolding in an old-fashioned configuration space, with wave functions
and other exotic Hilbert-space ingredients demoted from having an
ontological status. The complex numbers end up being necessary to
ensure that the Hilbert-space formalism is indeed a Markovian embedding.
\end{abstract}

\begin{center}
\global\long\def\quote#1{``#1"}%
\global\long\def\apostrophe{\textrm{'}}%
\global\long\def\slot{\phantom{x}}%
\global\long\def\eval#1{\left.#1\right\vert }%
\global\long\def\keyeq#1{\boxed{#1}}%
\global\long\def\importanteq#1{\boxed{\boxed{#1}}}%
\global\long\def\given{\vert}%
\global\long\def\mapping#1#2#3{#1:#2\to#3}%
\global\long\def\composition{\circ}%
\global\long\def\set#1{\left\{  #1\right\}  }%
\global\long\def\setindexed#1#2{\left\{  #1\right\}  _{#2}}%

\global\long\def\setbuild#1#2{\left\{  \left.\!#1\,\right|\,#2\right\}  }%
\global\long\def\complem{\mathrm{c}}%

\global\long\def\union{\cup}%
\global\long\def\intersection{\cap}%
\global\long\def\cartesianprod{\times}%
\global\long\def\disjointunion{\sqcup}%

\global\long\def\isomorphic{\cong}%

\global\long\def\setsize#1{\left|#1\right|}%
\global\long\def\defeq{\equiv}%
\global\long\def\conj{\ast}%
\global\long\def\overconj#1{\overline{#1}}%
\global\long\def\re{\mathrm{Re\,}}%
\global\long\def\im{\mathrm{Im\,}}%

\global\long\def\transp{\mathrm{T}}%
\global\long\def\tr{\mathrm{tr}}%
\global\long\def\adj{\dagger}%
\global\long\def\diag#1{\mathrm{diag}\left(#1\right)}%
\global\long\def\dotprod{\cdot}%
\global\long\def\crossprod{\times}%
\global\long\def\Probability#1{\mathrm{Prob}\left(#1\right)}%
\global\long\def\Amplitude#1{\mathrm{Amp}\left(#1\right)}%
\global\long\def\cov{\mathrm{cov}}%
\global\long\def\corr{\mathrm{corr}}%

\global\long\def\absval#1{\left\vert #1\right\vert }%
\global\long\def\expectval#1{\left\langle #1\right\rangle }%
\global\long\def\op#1{\hat{#1}}%

\global\long\def\bra#1{\left\langle #1\right|}%
\global\long\def\ket#1{\left|#1\right\rangle }%
\global\long\def\braket#1#2{\left\langle \left.\!#1\right|#2\right\rangle }%

\global\long\def\parens#1{(#1)}%
\global\long\def\bigparens#1{\big(#1\big)}%
\global\long\def\Bigparens#1{\Big(#1\Big)}%
\global\long\def\biggparens#1{\bigg(#1\bigg)}%
\global\long\def\Biggparens#1{\Bigg(#1\Bigg)}%
\global\long\def\bracks#1{[#1]}%
\global\long\def\bigbracks#1{\big[#1\big]}%
\global\long\def\Bigbracks#1{\Big[#1\Big]}%
\global\long\def\biggbracks#1{\bigg[#1\bigg]}%
\global\long\def\Biggbracks#1{\Bigg[#1\Bigg]}%
\global\long\def\curlies#1{\{#1\}}%
\global\long\def\bigcurlies#1{\big\{#1\big\}}%
\global\long\def\Bigcurlies#1{\Big\{#1\Big\}}%
\global\long\def\biggcurlies#1{\bigg\{#1\bigg\}}%
\global\long\def\Biggcurlies#1{\Bigg\{#1\Bigg\}}%
\global\long\def\verts#1{\vert#1\vert}%
\global\long\def\bigverts#1{\big\vert#1\big\vert}%
\global\long\def\Bigverts#1{\Big\vert#1\Big\vert}%
\global\long\def\biggverts#1{\bigg\vert#1\bigg\vert}%
\global\long\def\Biggverts#1{\Bigg\vert#1\Bigg\vert}%
\global\long\def\Verts#1{\Vert#1\Vert}%
\global\long\def\bigVerts#1{\big\Vert#1\big\Vert}%
\global\long\def\BigVerts#1{\Big\Vert#1\Big\Vert}%
\global\long\def\biggVerts#1{\bigg\Vert#1\bigg\Vert}%
\global\long\def\BiggVerts#1{\Bigg\Vert#1\Bigg\Vert}%
\global\long\def\ket#1{\vert#1\rangle}%
\global\long\def\bigket#1{\big\vert#1\big\rangle}%
\global\long\def\Bigket#1{\Big\vert#1\Big\rangle}%
\global\long\def\biggket#1{\bigg\vert#1\bigg\rangle}%
\global\long\def\Biggket#1{\Bigg\vert#1\Bigg\rangle}%
\global\long\def\bra#1{\langle#1\vert}%
\global\long\def\bigbra#1{\big\langle#1\big\vert}%
\global\long\def\Bigbra#1{\Big\langle#1\Big\vert}%
\global\long\def\biggbra#1{\bigg\langle#1\bigg\vert}%
\global\long\def\Biggbra#1{\Bigg\langle#1\Bigg\vert}%
\global\long\def\braket#1#2{\langle#1\vert#2\rangle}%
\global\long\def\bigbraket#1#2{\big\langle#1\big\vert#2\big\rangle}%
\global\long\def\Bigbraket#1#2{\Big\langle#1\Big\vert#2\Big\rangle}%
\global\long\def\biggbraket#1#2{\bigg\langle#1\bigg\vert#2\bigg\rangle}%
\global\long\def\Biggbraket#1#2{\Bigg\langle#1\Bigg\vert#2\Bigg\rangle}%
\global\long\def\angs#1{\langle#1\rangle}%
\global\long\def\bigangs#1{\big\langle#1\big\rangle}%
\global\long\def\Bigangs#1{\Big\langle#1\Big\rangle}%
\global\long\def\biggangs#1{\bigg\langle#1\bigg\rangle}%
\global\long\def\Biggangs#1{\Bigg\langle#1\Bigg\rangle}%

\global\long\def\vec#1{\mathbf{#1}}%
\global\long\def\vecgreek#1{\boldsymbol{#1}}%
\global\long\def\idmatrix{\mymathbb{1}}%
\global\long\def\projector{P}%
\global\long\def\permutationmatrix{\Sigma}%
\global\long\def\densitymatrix{\rho}%
\global\long\def\krausmatrix{K}%
\global\long\def\stochasticmatrix{\Gamma}%
\global\long\def\lindbladmatrix{L}%
\global\long\def\dynop{\Theta}%
\global\long\def\timeevop{U}%
\global\long\def\hadamardprod{\odot}%
\global\long\def\tensorprod{\otimes}%

\global\long\def\inprod#1#2{\left\langle #1,#2\right\rangle }%
\global\long\def\normket#1{\left\Vert #1\right\Vert }%
\global\long\def\hilbspace{\mathcal{H}}%
\global\long\def\samplespace{\Omega}%
\global\long\def\configspace{\mathcal{C}}%
\global\long\def\phasespace{\mathcal{P}}%
\global\long\def\spectrum{\sigma}%
\global\long\def\restrict#1#2{\left.#1\right\vert _{#2}}%
\global\long\def\from{\leftarrow}%
\global\long\def\statemap{\omega}%
\global\long\def\degangle#1{#1^{\circ}}%
\global\long\def\trivialvector{\tilde{v}}%
\global\long\def\eqsbrace#1{\left.#1\qquad\right\}  }%
\global\long\def\matf#1{\mathbf{#1}}%
\par\end{center}

\section{Introduction\label{sec:Introduction}}

One of the most mysterious features of quantum theory is its apparent
reliance on the complex numbers, which are included explicitly in
the textbook theory's Dirac\textendash von Neumann axioms (Dirac 1930,
von Neumann 1932)\nocite{Dirac:1930pofm,vonNeumann:1932mgdq}. There
are  arguments in the research literature that aim to establish the
necessity of the complex numbers in quantum theory through various
measurement scenarios (Renou et al. 2021)\nocite{RenouTrilloWeilenmannThinhTavakoliGisinAcinNavascues:1999qmoarhs}.
In this paper, I take a different and more foundational approach.

\section{The Complex Numbers\label{sec:The-Complex-Numbers}}

An algebra is a vector space, with its usual rules for addition and
multiplication by scalars, as well as an additive identity $0$, together
with a multiplication rule that takes two elements of the algebra
and returns another element of the algebra. A normed division algebra
has a norm $\verts q\geq0$ and includes for every nonzero element
$q\ne0$ a multiplicative reciprocal $1/q$, so that division is defined
for all nonzero divisors.

A key example is the set of real numbers $\mathbb{R}$, which form
a one-dimensional normed division algebra, with norm $\verts x\defeq\sqrt{x^{2}}$.

As follows from the Cayley\textendash Dickson construction (Dickson
1919)\nocite{Dickson:1919oqatgathotest} together with the Hurwitz
theorem (Hurwitz 1922)\nocite{Hurwitz:1922udkdqf}, there exists a
unique two-dimensional normed division algebra, called the complex
numbers $\mathbb{C}$. Elements of $\mathbb{C}$ can be identified
with ordered pairs $\left(a,b\right)$ of real numbers, $a$ and $b$,
together with the following arithmetic rules:
\begin{itemize}
\item Addition: $\left(a,b\right)+\left(c,d\right)\defeq\left(a+c,b+d\right)$.
\item Multiplication by scalars: $c\left(a,b\right)\defeq\left(ca,cb\right)$
for $c$ any real number.
\item Multiplication: $\left(a,b\right)\left(c,d\right)\defeq\left(ac-bd,ad+bc\right)$.
\end{itemize}
Notice that addition and multiplication by scalars are carried out
entry-wise, whereas multiplication of complex numbers with each other
takes a much less obvious form. To see why entry-wise multiplication
fails to give a division algebra, observe that $\left(a,0\right)$
is not the additive identity $\left(0,0\right)$, and yet, under entry-wise
multiplication, it lacks a multiplicative reciprocal, because then
$\left(1,1\right)$ is the multiplicative identity, and there are
no ordered pairs $\left(c,d\right)$ such that $\left(a,0\right)\left(c,d\right)=\left(1,1\right)$
under entry-wise multiplication.

The complex numbers also come with several other important operations: 
\begin{itemize}
\item Complex conjugation: $\left(a,b\right)^{\conj}\defeq\overconj{\left(a,b\right)}\defeq\left(a,-b\right)$.
\item Norm or modulus: $\verts{\left(a,b\right)}\defeq\sqrt{\left(a,b\right)^{\conj}\left(a,b\right)}=\sqrt{a^{2}+b^{2}}$.
\item Real part: $\re\left(a,b\right)\defeq a$.
\item Imaginary part: $\im\left(a,b\right)\defeq b$.
\end{itemize}

Any set of mathematical symbols that are in a one-to-one correspondence
with these ordered pairs and exhibit the same properties give a representation
of the complex numbers. For example, denote: 
\begin{equation}
\quote 1\defeq\left(1,0\right),\quad\quote i\defeq\left(0,1\right).\label{eq:Def1AndiForSimpleRepresentationComplexNumbers}
\end{equation}
 Then: 
\begin{equation}
z\defeq x+iy\defeq1x+iy=\left(1,0\right)x+\left(0,1\right)y=\left(x,y\right).\label{eq:DefGeneralComplexNumberFromOrderedPairs}
\end{equation}
 One has 
\begin{equation}
i^{2}=-1,\label{eq:iSquaredEqNegOne}
\end{equation}
 so one can formally write down the mystical-looking formula 
\begin{equation}
\quote{i=\sqrt{-1}.}\label{eq:iEqSqrtNegOne}
\end{equation}
 However, there is nothing particularly mysterious going on here.
Under the hood, there are just ordered pairs of real numbers with
a special multiplication rule.

From the Taylor series for the exponential function $e^{x}$, one
sees that 
\begin{align*}
e^{i\theta} & =1+i\theta+\frac{1}{2}\left(i\theta\right)^{2}+\frac{1}{6}\left(i\theta\right)^{3}+\cdots\\
 & =1+i\theta-\frac{1}{2}\theta^{2}-i\frac{1}{6}\theta^{3}\pm\cdots\\
 & =\left(1-\frac{1}{2}\theta^{2}\pm\cdots\right)+i\left(\theta-\frac{1}{6}\theta^{3}\pm\cdots\right).
\end{align*}
 Identifying the two quantities in parentheses with the respective
Taylor series for the cosine and sine functions, one ends up with
the Euler formula: 
\begin{equation}
e^{i\theta}=\cos\theta+i\sin\theta.\label{eq:EulerFormula}
\end{equation}
 As a special case, for $\theta=\pi$, this result leads to the famous
Euler identity, 
\begin{equation}
e^{i\pi}+1=0,\label{eq:EulerIdentity}
\end{equation}
 which unites the fundamental constants $0$, $1$, $e$, $i$, and
$\pi$ together with the basic mathematical operations of addition
$+$ and equality $=$.

From trigonometry, for a right triangle with opening angle $\theta$,
adjacent side of length $x$, opposite side of length $y$, and hypotenuse
of length $r$, one has the basic relationships 
\begin{equation}
\begin{aligned}x & =r\cos\theta,\\
y & =r\sin\theta,\\
r & =\sqrt{x^{2}+y^{2}}.
\end{aligned}
\label{eq:TrigRelationshipsForComplexNumber}
\end{equation}
 These formulas give the polar form of a complex number $z$: 
\begin{equation}
z=x+iy=r\left(\cos\theta+i\sin\theta\right)=re^{i\theta}.\label{eq:PolarFormComplexNumber}
\end{equation}
 Here $r=\verts z$ is the norm or modulus of the complex number,
and $e^{i\theta}$ is called its phase factor, or phase for short.
Phases have unit modulus: 
\begin{equation}
\verts{e^{i\theta}}=1.\label{eq:PhaseUnitMod}
\end{equation}
 Taking the modulus of a general complex number $z=re^{i\theta}$
erases its phase information $e^{i\theta}$: 
\begin{equation}
\verts z=\verts{re^{i\theta}}=r.\label{eq:ModGeneralComplexNumberPolar}
\end{equation}

There exist other representations of the complex numbers $\mathbb{C}$.
One interesting example is the \textquoteleft real\textquoteright{}
$2\times2$ representation (see, for example, Myrheim 1999)\nocite{Myrheim:1999qmoarhs},
in which one introduces the following pair of $2\times2$ matrices,
both of which have only real-valued entries: 
\begin{equation}
\matf 1\defeq\begin{pmatrix}1 & 0\\
0 & 1
\end{pmatrix},\quad\matf I\defeq\begin{pmatrix}0 & -1\\
1 & 0
\end{pmatrix}.\label{eq:Real2by2MatricesFor1Andi}
\end{equation}
 These matrices satisfy 
\begin{equation}
\matf I^{2}=-\matf 1.\label{eq:Real2by2iSqEqNegOne}
\end{equation}
 It follows that any linear combination of the two matrices $\matf 1$
and $\matf I$ with real-valued coefficients $x$ and $y$ defines
a $2\times2$ matrix of the specific form 
\begin{equation}
\matf Z\defeq\matf 1x+\matf Iy=\begin{pmatrix}x & -y\\
y & x
\end{pmatrix}.\label{eq:Real2by2ComplexNumberAsMatrix}
\end{equation}
 One can check that matrices of this form have  the same arithmetic
behavior as complex numbers, so they indeed provide an alternative
 representation of $\mathbb{C}$. Moreover, taking the transpose
of $\matf Z$  implements complex conjugation, 
\begin{equation}
\matf Z^{\transp}=\matf 1x-\matf Iy,\label{eq:Real2by2ComplexNumberTransposeAsConjugate}
\end{equation}
 taking the trace gives twice the real part, 
\begin{equation}
\tr\left(\matf Z\right)=2x,\label{eq:Real2by2ComplexNumberTraceAsTwiceRealPart}
\end{equation}
 and taking the determinant yields the modulus-squared, 
\begin{equation}
\det\left(\matf Z\right)=x^{2}+y^{2}.\label{eq:Real2by2ComplexNumberDeterminantAsModSquared}
\end{equation}
 Additionally, recall that a $2\times2$ matrix that implements rotations
by an angle $\theta$ is given by 
\begin{equation}
\operatorname{Rot}\left(\theta\right)=\begin{pmatrix}\cos\theta & -\sin\theta\\
\sin\theta & \cos\theta
\end{pmatrix},\label{eq:2by2RotationMatrix}
\end{equation}
and observe that one can decompose the right-hand side as 
\[
\begin{pmatrix}\cos\theta & -\sin\theta\\
\sin\theta & \cos\theta
\end{pmatrix}=\begin{pmatrix}1 & 0\\
0 & 1
\end{pmatrix}\cos\theta+\begin{pmatrix}0 & -1\\
1 & 0
\end{pmatrix}\sin\theta.
\]
 That is, 
\begin{equation}
\operatorname{Rot}\left(\theta\right)=\matf 1\cos\theta+\matf I\sin\theta=e^{\matf I\theta},\label{eq:2by2RotationMatrixAsReal2by2ComplexNumber}
\end{equation}
 where the exponential of a matrix is  defined by its Taylor expansion.
This result is just the \textquoteleft real\textquoteright{} $2\times2$
representation of the Euler formula \eqref{eq:EulerFormula}.

The 2 \texttimes{} 2 matrix $\matf I$ here is called a linear complex
structure. Notice that it does not involve any appearances of a \textquoteleft literal\textquoteright{}
$i$, but only real numbers. However, the complex numbers, as an algebraic
structure, are still present. The complex numbers are fundamentally
an algebraic structure, so any algebraic structure isomorphic to $\mathbb{C}$
is really just $\mathbb{C}$ itself.\footnote{By contrast, Renou et al. (2021)\nocite{RenouTrilloWeilenmannThinhTavakoliGisinAcinNavascues:1999qmoarhs}
argue that replacing $i$ with the $2\times2$ matrix $\matf I$ qualifies
as removing the complex numbers from quantum theory and results in
a truly real theory. This argument is not correct\textemdash their
theory still involves the complex numbers because their theory still
involves factors of $\matf I$, with all its usual algebraic properties.
Moreover, when replacing factors of $i$ with $\matf I$, it follows
from straightforward mathematical identities that one must give up
certain assumptions about tensor-factorizability in order to obtain
an equivalent formulation of standard quantum theory. Renou et al.
(2021) decide not to give up those assumptions, and, rather unsurprisingly,
they ultimately find that their still-complex theory disagrees with
some of the predictions of standard quantum theory. It is not clear
how significant it is to demonstrate the inequivalence between standard
quantum theory and an arbitrarily defined new theory that still involves
the complex numbers.}

Returning to a more abstract representation of complex numbers as
expressions of the form $z=x+iy$, it will turn out to be useful to
introduce an operator $K$ that implements complex conjugation: 
\begin{equation}
Kz\defeq\overconj zK.\label{eq:DefComplexConjOperator}
\end{equation}
 By construction, $K$ anticommutes with $i$: 
\begin{equation}
Ki=-iK.\label{eq:ComplexConjOpAnticommutesWithi}
\end{equation}
 It follows that the symbols $i$, $K$, and $iK$ generate a Clifford
algebra known as the pseudo-quaternions (Stueckelberg 1960)\nocite{Stueckelberg:1960qtirhs}:
\begin{equation}
-i^{2}=K^{2}=\left(iK\right)^{2}=\left(i\right)\left(K\right)\left(iK\right)=1.\label{eq:ElementaryPseudoQuaternionsAlgebra}
\end{equation}
 In the \textquoteleft real\textquoteright{} $2\times2$ matrix representation
\eqref{eq:Real2by2ComplexNumberAsMatrix}, $K$ is represented by
a $2\times2$ matrix with real-valued entries that happens to coincide
with the first Pauli sigma matrix $\sigma_{x}$: 
\begin{equation}
\matf K\defeq\begin{pmatrix}0 & 1\\
1 & 0
\end{pmatrix}=\sigma_{x}.\label{eq:ComplexConjOpAsPauliSigmaMatrix}
\end{equation}

\section{Markovian Embeddings\label{sec:Markovian-Embeddings}}

A Markov process is a dynamical system for which the laws that pick
out future configurations, either deterministically or probabilistically,
depend only on the system's present configuration and not on prior
configurations in the past. It will be fruitful to examine the ways
in which a non-Markovian system can be made to look Markovian by a
suitable change of  representation. 

As a simple first example, consider a \emph{discrete} system, whose
kinematics consists of configurations 
\begin{equation}
x=1,2,3,\dots,N\label{eq:2ndOrderDiscreteSystemConfigurations}
\end{equation}
 that make up a configuration space $\configspace$. At the level
of dynamics, suppose that the system has a second-order deterministic
law 
\begin{equation}
x\left(t+1\right)=F\left(x\left(t\right),x\left(t-1\right)\right),\label{eq:2ndOrderDiscreteSystemLaw}
\end{equation}
 where $F$ is some given function with two arguments. This dynamical
law is said to be non-Markovian of second order because picking out
the system's configuration at a later time $t+1$ requires specifying
the system's configuration not merely at the present time $t$, but
also at the previous time $t-1$. This model as a whole can be understood
as a discrete version of a simple system of one degree of freedom
in Newtonian mechanics.

One can then always make the following nomology\textendash ontology
trade-off: Simplify the model's dynamical law, or nomology, to be
first order, or Markovian, at the cost of complicating the model's
configuration space, meaning its ontology. According to the new kinematics,
the \textquoteleft states\textquoteright{} are now ordered pairs $\left(x,y\right)$
making up a \textquoteleft state space\textquoteright{} $\configspace\cartesianprod\configspace$,
and the new dynamics is given by the \emph{pair} of \emph{first-order}
equations 
\begin{equation}
\begin{aligned}x\left(t+1\right) & =F\left(x\left(t\right),y\left(t\right)\right),\\
y\left(t+1\right) & =x\left(t\right).
\end{aligned}
\label{eq:FirstOrderMarkovianEmbeddingDiscrete}
\end{equation}
 One therefore ends up with a Markovian embedding of the originally
second-order, non-Markovian model.

This Markovian embedding also features an emergent set of transformations
that are allowed to mix up the variables $x$ and $y$ for convenience
as long as the transformations do not entail losing any information.
For example, one can define $X\defeq x+y$ and $Y\defeq x-y$, which
would then satisfy the equations 
\begin{align}
X\left(t+1\right) & =F\left(\frac{X\left(t\right)+Y\left(t\right)}{2},\frac{X\left(t\right)-Y\left(t\right)}{2}\right)+\frac{X\left(t\right)+Y\left(t\right)}{2},\label{eq:FirstOrderMarkovianEmbeddingDiscreteAltVariables}\\
Y\left(t+1\right) & =F\left(\frac{X\left(t\right)+Y\left(t\right)}{2},\frac{X\left(t\right)-Y\left(t\right)}{2}\right)-\frac{X\left(t\right)+Y\left(t\right)}{2}.\nonumber 
\end{align}
 After solving these alternative first-order equations, one could
then immediately write down the solutions to the original equations
as $x\left(t\right)=\left(X\left(t\right)+Y\left(t\right)\right)/2$
and $y\left(t\right)=\left(X\left(t\right)-Y\left(t\right)\right)/2$.

As a second example, consider a \emph{continuous} system whose kinematics
consists of configurations $x$ making up a configuration space $\configspace$,
with dynamics given by a second-order deterministic law 
\begin{equation}
\ddot{x}\left(t\right)=F\left(x\left(t\right),\dot{x}\left(t\right)\right),\label{eq:2ndOrderContinuousSystemLaw}
\end{equation}
 using Newton's fluxion notation in which dots denote time derivatives.
Again, one can make a nomology\textendash ontology trade-off, with
the new kinematics consisting of \textquoteleft states\textquoteright{}
$\left(x,y\right)$ making up a \textquoteleft state space\textquoteright{}
$\configspace\cartesianprod\configspace$ and the new dynamics consisting
of a Markovian embedding given by the pair of first-order equations
\begin{equation}
\begin{aligned}\dot{x}\left(t\right) & =y\left(t\right),\\
\dot{y}\left(t\right) & =F\left(x\left(t\right),y\left(t\right)\right).
\end{aligned}
\label{eq:1stOrderMarkovianEmbeddingContinuous}
\end{equation}
 As before, this Markovian embedding features an emergent set of transformations
that can freely mix up $x$ and $y$ as long as they do not lose track
of any information. This example is a template for the Hamiltonian
formulation of Newtonian mechanics, with $x$ playing the role of
a canonical coordinate, $y$ playing the role of a canonical momentum,
and the set of transformations mixing up $x$ and $y$ containing
the set of canonical transformations.

For $t_{0}$ any choice of origin for the time coordinate, one can
define a time-reversal transformation by the replacement 
\begin{equation}
t_{0}+t\mapsto t_{0}-t,\label{eq:TimeReversalTransf}
\end{equation}
 so that translations forward in time from $t_{0}$ become translations
backward in time from $t_{0}$, and vice versa. For simplicity, the
origin for the time coordinate will be taken to be $t_{0}=0$. Then
it follows from the chain rule that for the original second-order
formulation of the continuous model, time-reversal transformations
act on the degree of freedom $x\left(t\right)$ and its time derivatives
according to the rules 
\begin{align}
x\left(t\right) & \mapsto+x\left(-t\right),\nonumber \\
\dot{x}\left(t\right) & \mapsto-\dot{x}\left(-t\right),\label{eq:ContinuousModelTimeReversalDof}\\
\ddot{x}\left(t\right) & \mapsto+\ddot{x}\left(-t\right).\nonumber 
\end{align}
 The original second-order law \eqref{eq:2ndOrderContinuousSystemLaw}
is therefore time-reversal invariant precisely if 
\begin{equation}
F\left(x,-\dot{x}\right)=F\left(x,\dot{x}\right).\label{eq:2ndOrderContinuousTimeReversalInvariantLaw}
\end{equation}

For the equivalent Markovian embedding, a time-reversal transformation
must instead be required to take the form 
\begin{align}
x\left(t\right) & \mapsto+x\left(-t\right),\nonumber \\
\dot{x}\left(t\right) & \mapsto-\dot{x}\left(-t\right),\label{eq:1stOrderContinuousMarkovianEmbeddingTimeReversal}\\
y\left(t\right) & \mapsto-y\left(-t\right),\nonumber \\
\dot{y}\left(t\right) & \mapsto+\dot{y}\left(-t\right),\nonumber 
\end{align}
 where $y\left(t\right)$ and $\dot{y}\left(t\right)$ transform with
opposite signs as compared with $x\left(t\right)$ and $\dot{x}\left(t\right)$.
The first-order Markovian laws then transform as 
\begin{align}
\dot{x}\left(t\right)=y\left(t\right) & \mapsto-\dot{x}\left(t\right)=-y\left(t\right),\nonumber \\
\dot{y}\left(t\right)=F\left(x\left(t\right),y\left(t\right)\right) & \mapsto\dot{y}\left(t\right)=F\left(x\left(t\right),-y\left(t\right)\right).\label{eq:1stOrderContinuousMarkovianEmbeddingTimeRevLaws}
\end{align}
 The dynamical laws are time-reversal-invariant if 
\begin{equation}
F\left(x,-y\right)=F\left(x,y\right).\label{eq:1stOrderContinuousMarkovianEmbeddingTimeRevInvariantLaw}
\end{equation}

Many Markovian embeddings are naturally expressible using complex
numbers. For the present example, one can define 
\begin{equation}
z\left(t\right)\defeq x\left(t\right)+iy\left(t\right)\label{eq:1stOrderContinuousMarkovianEmbeddingComplexVar}
\end{equation}
 and 
\begin{equation}
\mathcal{F}\bigparens{z\left(t\right),z^{\conj}\left(t\right)}\defeq y\left(t\right)+iF\left(x\left(t\right),y\left(t\right)\right).\label{eq:1stOrderContinuousMarkovianEmbeddingComplexLaw}
\end{equation}
 Then one ends up with a \emph{single} first-order Markovian law:
\begin{equation}
\dot{z}\left(t\right)=\mathcal{F}\bigparens{z\left(t\right),z^{\conj}\left(t\right)}.\label{eq:1stOrderContinuousMarkovianEmbeddingComplexEquations}
\end{equation}
 So one has made the dynamical law look even simpler, at the cost
of making the ontology murkier and introducing the purportedly exotic
complex numbers into the story.

This complex form of the dynamics makes it particularly easy to implement
time-reversal transformations: 
\begin{equation}
\begin{aligned}z\left(t\right) & \mapsto Kz\left(-t\right),\\
\dot{z}\left(t\right) & \mapsto-K\dot{z}\left(-t\right).
\end{aligned}
\label{eq:1stOrderContinuousMarkovianEmbeddingComplexTimeRevFromOp}
\end{equation}
 Notice, in particular, how the transformation law for $z\left(t\right)$
entails \emph{both} a change in its argument from $t$ to $-t$ \emph{together}
with an application of the complex-conjugation operator $K$, as originally
introduced in \eqref{eq:DefComplexConjOperator}. Meanwhile, the first-order
Markovian law \eqref{eq:1stOrderContinuousMarkovianEmbeddingComplexEquations}
transforms according to: 
\begin{equation}
-K\dot{z}\left(-t\right)=\mathcal{F}\bigparens{z^{\conj}\left(-t\right),z\left(-t\right)}.\label{eq:1stOrderContinuousMarkovianEmbeddingComplexEquationsTimeRev}
\end{equation}
 The dynamical law is therefore time-reversal-invariant if 
\begin{equation}
-K\mathcal{F}\left(z^{\conj},z\right)=\mathcal{F}\left(z,z^{\conj}\right).\label{eq:1stOrderContinuousMarkovianEmbeddingComplexTimeRevInvariant}
\end{equation}

This trick of embedding a non-Markovian model into a Markovian model
with a more complicated state space generalizes to systems with 
\begin{itemize}
\item arbitrarily many degrees of freedom,
\item non-Markovian dynamics of arbitrarily high order, and
\item stochastic dynamical laws.
\end{itemize}
The embedded-Markovian formulation may have a murky ontology; a large
class of strange-looking transformations of its state-space variables;
and exotic mathematical structures, such as the complex numbers. If
one were handed such an embedded-Markovian formulation at the outset,
one could easily imagine endless debates over its correct physical
interpretation.

\section{The Strocchi\textendash Heslot Formulation\label{sec:The-Strocchi-Heslot-Formulation}}

A closed quantum system with state vector $\ket{\Psi\left(t\right)}$
evolving according to the \emph{Schrödinger equation} 
\begin{equation}
i\frac{\partial\ket{\Psi\left(t\right)}}{\partial t}=H\ket{\Psi\left(t\right)}\label{eq:SchroEq}
\end{equation}
 (conventionally written with a partial time derivative $\partial/\partial t$)
is a Markov process with a murky ontology, features a large class
of strange (unitary change-of-basis) transformations, and involves
the complex numbers. Reversing the logic of the previous section,
one might then naturally ask whether there is some non-Markovian system
with a more transparent ontology for which this Hilbert-space formulation
is merely its Markovian embedding. Remarkably, the answer turns out
to be in the affirmative. One can find hints in this direction from
works like that of Glick and Adami (2020)\nocite{GlickAdami:2020manmqm}
that demonstrate clear instances of non-Markovian behavior for reasonably
straightforward, \emph{closed} quantum systems. (For \emph{open} systems,
non-Markovianity has long been known to be ubiquitous, in both the
classical and quantum cases, as a result of prosaic feedback effects
from the environment.)

Moreover, there exists an intriguing but little-known reformulation
of quantum theory as a \emph{classical} Hamiltonian system of coupled
harmonic oscillators. This reformulation comes from two beautiful
papers, the first by Franco Strocchi (1966)\nocite{Strocchi:1966ccqm}
and the second by André Heslot (1985)\nocite{Heslot:1985qmaact}.\footnote{For the generalization to infinite-dimensional Hilbert spaces, see
the work of Ashtekar and Schilling (1999)\nocite{AshtekarSchilling:1999gfoqm}.}

To start, consider a quantum system with an $N$-dimensional Hilbert
space $\hilbspace$, a state vector $\ket{\Psi\left(t\right)}$, and
a Hamiltonian $H=H^{\adj}$. Again, the Schrödinger equation reads
\[
i\frac{\partial\ket{\Psi\left(t\right)}}{\partial t}=H\ket{\Psi\left(t\right)}.
\]
 Next, pick an arbitrary orthonormal basis\textemdash a choice to
be revisited shortly\textemdash so that the state vector $\ket{\Psi\left(t\right)}$
is represented by the column matrix 
\begin{equation}
\begin{pmatrix}\Psi_{1}\left(t\right)\\
\vdots\\
\Psi_{N}\left(t\right)
\end{pmatrix},\label{eq:StateVectorAsColumnMatrix}
\end{equation}
 and the Hamiltonian $H$ is represented by the square matrix 
\begin{equation}
\begin{pmatrix}H_{11} & H_{12}\\
H_{21} & \ddots\\
 &  & H_{NN}
\end{pmatrix}.\label{eq:HamiltonianAsSquareMatrix}
\end{equation}
 Writing out the Schrödinger equation in terms of individual entries,
one has 
\begin{equation}
i\frac{d\Psi_{i}\left(t\right)}{dt}=\sum^{N}_{j=1}H_{ij}\Psi_{j}\left(t\right).\label{eq:SchroEqInComponents}
\end{equation}
 Then decompose all the individual entries of $\ket{\Psi\left(t\right)}$
and $H$ into their real and imaginary parts:\footnote{The $1/\sqrt{2}$ normalization is convenient for rewriting Poisson
brackets in terms of complex variables.} 
\begin{equation}
\begin{aligned}\Psi_{i}\left(t\right) & =\frac{1}{\sqrt{2}}\left(q_{i}\left(t\right)+ip_{i}\left(t\right)\right),\\
H_{ij} & =A_{ij}+iB_{ij},
\end{aligned}
\label{eq:StateVectorAndHamiltonianInComponents}
\end{equation}
 where $H=H^{\adj}$ implies the symmetry conditions 
\begin{equation}
\begin{aligned}A_{ij} & =A_{ji},\\
B_{ij} & =-B_{ji}.
\end{aligned}
\label{eq:HamiltonianEntriesRealityConditions}
\end{equation}

The Schrödinger equation then becomes 
\begin{equation}
i\frac{d\left(q_{i}\left(t\right)+ip_{i}\left(t\right)\right)}{dt}=\sum^{N}_{j=1}\left(A_{ij}+iB_{ij}\right)\left(q_{j}\left(t\right)+ip_{j}\left(t\right)\right).\label{eq:SchroEqRealComponentsIntermed}
\end{equation}
 Breaking up this equation into its real and imaginary parts, and
simplifying fully, one ends up with a system of $2N$ classical Hamiltonian
equations of motion: 
\begin{equation}
\begin{aligned}\dot{q}_{i}\left(t\right) & =\frac{\partial H_{\textrm{SH}}}{\partial p_{i}},\\
\dot{p}_{i}\left(t\right) & =-\frac{\partial H_{\textrm{SH}}}{\partial q_{i}}.
\end{aligned}
\label{eq:StrocchiHeslotHamiltonEqs}
\end{equation}
 Here, $H_{\textrm{SH}}$ is the Strocchi\textendash Heslot Hamiltonian,
which is just a classical Hamiltonian that ends up with a manifestly
bilinear form as a result of the linearity of the Schrödinger equation:
\begin{equation}
H_{\textrm{SH}}\defeq\sum_{k,l}\left(\frac{1}{2}A_{kl}p_{k}p_{l}+B_{kl}q_{k}p_{l}+\frac{1}{2}A_{kl}q_{k}q_{l}\right).\label{eq:StrocchiHeslotClassicalHamiltonian}
\end{equation}

Schrödinger time evolution is therefore isomorphic, on the Strocchi\textendash Heslot
formulation, to the dynamics of a system of \emph{classical} coupled
harmonic oscillators. Changing orthonormal basis on the Hilbert-space
side is then equivalent to carrying out a linear canonical transformation
on the Strocchi\textendash Heslot side.

Moreover, under the most straightforward notion of a time-reversal
transformation, one has $q_{i}\left(t\right)\mapsto q_{i}\left(-t\right)$
and $p_{i}\mapsto-p_{i}\left(-t\right)$, with an extra minus sign
for $p_{i}\left(t\right)$, precisely as in \eqref{eq:1stOrderContinuousMarkovianEmbeddingTimeReversal}.
Hence, when $q_{i}\left(t\right)$ and $p_{i}\left(t\right)$ are
recombined into the components $\Psi_{i}\left(t\right)=\parens{1/\sqrt{2}}\left(q_{i}\left(t\right)+ip_{i}\left(t\right)\right)$
of the system's state vector, one sees in analogy with the transformation
law for $z\left(t\right)$ in \eqref{eq:1stOrderContinuousMarkovianEmbeddingComplexTimeRevFromOp}
that $\Psi_{i}\left(t\right)$ transforms according to the rule 
\begin{equation}
\Psi_{i}\left(t\right)\mapsto K\Psi_{i}\left(-t\right).\label{eq:TimeReversalStateVectorComponents}
\end{equation}
 That is, $\Psi_{i}\left(t\right)$ experiences\emph{ both} a reversal
of the time argument from $t$ to $-t$ \emph{together} with an application
of the complex-conjugation operator $K$. The Strocchi\textendash Heslot
formulation therefore gives a nice way of seeing why state vectors
obey this odd-looking transformation rule.\footnote{If one were to allow for a linear canonical transformation as part
of the time-reversal transformation, then $\Psi_{i}\left(t\right)$
would instead transform as $\Psi_{i}\left(t\right)\mapsto VK\Psi_{i}\left(-t\right)$,
where $V$ is some unitary matrix. This more generic transformation
law is, in fact, the most general textbook rule for the time reversal
of a state vector.}

As a particularly interesting choice of orthonormal basis, picking
the energy eigenbasis corresponds to carrying out a linear canonical
transformation to the set of normal modes, for which the oscillators
are uncoupled and independent, with individual frequencies proportional
to the eigenvalues of the original quantum Hamiltonian $H$. Mathematically
speaking, this canonical frame is arguably the simplest description
of the evolution for any quantum system\textemdash after all, there
are few systems simpler than a set of uncoupled, classical harmonic
oscillators.

Some interpretations of quantum theory attempt to reify its usual
Markovian formulation. For such interpretations, it is not immediately
clear why the Hilbert-space formulation should be favored, metaphysically
speaking, over the Strocchi\textendash Heslot formulation. If, however,
the Strocchi\textendash Heslot formulation is reified, then is the
implication that the universe is just a collection of harmonic oscillators?
In that case, why should having zero amplitude\textemdash say, for
a \textquoteleft branch\textquoteright{} or \textquoteleft world\textquoteright{}
in some versions of Everettian quantum theory\textemdash be taken
to mean a lack of physical existence? After all, a harmonic oscillator
that is not oscillating still physically exists. One therefore sees
the dangers of trying to build an ontology directly out of a purely
mathematical structure, simply because mathematical structures have
many different-looking formulations, and what fixes the formulation
to be one over another?

Notice that for a quantum system with an $N$-dimensional Hilbert
space, for $N$ finite, the corresponding Strocchi\textendash Heslot
formulation features a continuously infinite state space of dimension
$2N$. In particular, the Strocchi\textendash Heslot state space is
continuously infinite even for the simplest possible quantum system:
a qubit, which has $N=2$. Hence, if the Strocchi\textendash Heslot
state space can be understood as the state space of a Markovian embedding
of some underlying non-Markovian system with a discrete configuration
space consisting of just $N$ elements, then the dynamics of that
underlying system is presumably going to be extremely non-Markovian,
far more so than second order.

\section{Indivisible Stochastic Processes\label{sec:Indivisible-Stochastic-Processes}}

An indivisible stochastic process is a simple generalization of a
non-Markovian stochastic process. The new axioms, which are considerably
simpler than the Dirac\textendash von Neumann axioms and do not involve
Hilbert spaces or even the complex numbers, are a combination of fixed
and contingent ingredients. The fixed ingredients, which are the same
in every real-world instantiation or run of the given indivisible
model, are as follows: 
\begin{itemize}
\item a kinematics consisting of a configuration space $\configspace$,
and
\item a dynamics consisting of transition maps $\mapping{\stochasticmatrix\left(t\from t_{0}\right)}{\configspace\cartesianprod\configspace}{\left[0,1\right]}$
for pairs of times $t,t_{0}$ taken from respective index sets $\mathcal{T}$,
called the target times, and $\mathcal{T}_{0}\subset\mathcal{T}$,
called the conditioning times, where each transition map $\stochasticmatrix\left(t\from t_{0}\right)$
encodes first-order conditional transition probabilities connecting
$t_{0}$ to $t$.
\end{itemize}
The contingent ingredient, which will generically differ between real-world
instantiations or runs of the model, is 
\begin{itemize}
\item a standalone probability distribution $\mapping{p\left(t\right)}{\configspace}{\left[0,1\right]}$
for times $t$ in the set of target times $\mathcal{T}$.
\end{itemize}
Labeling distinct configurations in $\configspace$ with an index
$i$ or $j$, the standard rules of ordinary probability theory include
the usual law of total probability, which implies that there is a
\emph{linear} marginalization rule connecting standalone probabilities
$p_{j}\left(t_{0}\right)$ at conditioning times $t_{0}$ and standalone
probabilities $p_{i}\left(t\right)$ at target times $t$: 
\begin{equation}
p_{i}\left(t\right)=\sum_{j}\stochasticmatrix_{ij}\left(t\from t_{0}\right)p_{j}\left(t_{0}\right).\label{eq:LawOfTotalProbabilityFromTransitionMatrix}
\end{equation}
 Here, the two-indexed quantities $\stochasticmatrix_{ij}\left(t\from t_{0}\right)$
define a transition matrix, which is then a (column) stochastic matrix,
meaning a matrix whose entries are nonnegative real numbers and each
of whose columns sum to $1$.

Indivisibility is a simple and remarkably new idea. One can think
of it as \textquoteleft a failure of iterativeness.\textquoteright{}
The conceptual foundations of indivisibility originated in the theory
of quantum channels (Wolf, Cirac 2008)\nocite{WolfCirac:2008dqc},
but they were only first applied to ordinary probability theory and
stochastic processes within the last few years (Milz, Modi 2021)\nocite{MilzModi:2021qspaqnp}.

In short, for an indivisible stochastic process, not all target times
are valid conditioning times at the level of the transition matrices.
That is, even if the transition matrices $\stochasticmatrix\left(t\from t_{0}\right)$
and $\stochasticmatrix\left(t^{\prime}\from t_{0}\right)$ exist for
times $t>t^{\prime}>t_{0}$, where $t^{\prime}$ is intermediate between
$t_{0}$ and $t$, it may be the case that 
\begin{equation}
\stochasticmatrix\left(t\from t_{0}\right)\ne\stochasticmatrix\left(t\from t^{\prime}\right)\stochasticmatrix\left(t^{\prime}\from t_{0}\right)\label{eq:Indivisibility}
\end{equation}
 for all matrices $\stochasticmatrix\left(t\from t^{\prime}\right)$
that satisfy the conditions of being stochastic and that are properly
interpreted by the theory as describing first-order conditional transition
probabilities connecting $t^{\prime}$ to $t$. The inequality \eqref{eq:Indivisibility}
is called indivisibility.

As emphasized, for instance, in work by Gillespie (1998, 2000)\nocite{Gillespie:1998dtsoasp,Gillespie:2000nsp},
a traditional non-Markovian stochastic process features an infinite
hierarchy or Kolmogorov tower of conditional probabilities of first
order, second order, third order, and higher orders: 
\begin{equation}
\begin{aligned} & p\left(i,t\given j_{1},t_{1}\right),\\
 & p\left(i,t\given j_{1},t_{1};j_{2},t_{2}\right),\\
 & p\left(i,t\given j_{1},t_{1};j_{2},t_{2};j_{3},t_{3}\right).
\end{aligned}
\label{eq:NonMarkovianTowerConditionalProbabilities}
\end{equation}
 From this Kolmogorov tower of conditional probabilities, together
with standalone probabilities $p\left(i,t\right)$, one can go on
to construct all possible two-way joint probabilities, three-way joint
probabilities, and so on.\footnote{For pedagogical treatments of the theory of stochastic processes,
see Rosenblatt (1962), Parzen (1962), Doob (1990), or Ross (1995)\nocite{Rosenblatt:1962rp,Parzen:1962sp,Doob:1990sp,Ross:1995sp}.}

In a sense, an indivisible stochastic process represents an entire
\emph{equivalence class} of non-Markovian stochastic processes of
arbitrary order that agree on the sparse set of first-order conditional
probabilities encoded in the transition maps $\stochasticmatrix\left(t\from t_{0}\right)$.
Each such non-Markovian stochastic process in this equivalence class
is compatible with the laws of the given indivisible stochastic process,
and is called a non-Markovian realizer.\footnote{The author thanks Alexander Meehan for suggesting this terminology.}
By contrast, a Markov process entails making the approximation that
all second- and higher-order conditional probabilities $p\left(i,t\given j_{1},t_{1};j_{2},t_{2};\dots\right)$
are fixed and numerically equal to first-order conditional probabilities,
conditioned on the time closest to the target time $t$.

To establish the equivalence between indivisible stochastic processes
and quantum systems, called the stochastic\textendash quantum correspondence,
one needs to define the stochastic-to-quantum direction of the equivalence,
and then the quantum-to-stochastic direction.

For the stochastic-to-quantum direction, one starts by considering
any indivisible stochastic process with $N$ configurations and then
picking a conditioning time, taken to be $0$ without loss of generality,
as well as a target time $t$. Then, given the transition matrix $\stochasticmatrix\left(t\from0\right)$,
one proceeds to \textquoteleft solve\textquoteright{} the nonnegativity
condition $\stochasticmatrix_{ij}\left(t\from0\right)\geq0$ by introducing
a nonunique \textquoteleft potential\textquoteright{} matrix $\dynop\left(t\from0\right)$
whose entries are allowed to be complex-valued, according to the entry-by-entry
equality 
\begin{equation}
\stochasticmatrix_{ij}\left(t\from0\right)=\verts{\dynop_{ij}\left(t\from0\right)}^{2}.\label{eq:TransitionMatrixFroModSquarePotentialMatrix}
\end{equation}

If it happens that the transition matrix $\stochasticmatrix\left(t\from0\right)$
is unistochastic (Stueckelberg 1952; Horn 1954; Thompson 1989; Nylen,
Tam 1993)\nocite{Stueckelberg:1952theuds,Horn:1954dsmatdoarm,Thompson:1989uln,NylenTamUhlig:1993oteopsonhasm},
then $\dynop\left(t\from0\right)$ can be assumed to be a unitary
matrix $\timeevop\left(t\from0\right)$, satisfying 
\begin{equation}
\timeevop^{\adj}\left(t\from0\right)=\timeevop^{-1}\left(t\from0\right)\label{eq:TimeEvOpUnitary}
\end{equation}
 and called the system's time-evolution operator. 

Otherwise, one can define a set of $N\times N$ matrices $\krausmatrix_{\beta}\left(t\from0\right)$,
with $\beta=1,\dots,N$, that each take one column from $\dynop\left(t\from0\right)$
and are otherwise filled with $0$s. It is a straightforward exercise
to show that the matrices $\krausmatrix_{\beta}\left(t\from0\right)$
are Kraus operators, meaning that they satisfy the Kraus identity
\begin{equation}
\sum_{\beta}\krausmatrix^{\adj}_{\beta}\left(t\from0\right)\krausmatrix_{\beta}\left(t\from0\right)=\idmatrix,\label{eq:KrausIdentity}
\end{equation}
 where $\idmatrix$ is the identity matrix, and also that the transition
matrix $\stochasticmatrix\left(t\from0\right)$ has the following
Kraus decomposition: 
\begin{equation}
\stochasticmatrix_{ij}\left(t\from0\right)=\sum^{N}_{\beta=1}\verts{\krausmatrix_{\beta,ij}\left(t\from0\right)}^{2}.\label{eq:TransitionMatrixFromKrausDecomposition}
\end{equation}
 It follows from an application of the Stinespring dilation theorem
(Stinespring 1955; Barandes 2025a, 2025b, 2026)\nocite{Stinespring:1955pfoc,Barandes:2025tsqc,Barandes:2025qsaisp,Barandes:2023tsqt}
that by increasing $N$ to $N^{\prime}$, which is bounded from above
by $N^{3}$, one can replace the $N\times N$ matrix $\dynop\left(t\from0\right)$
with an $N^{\prime}\times N^{\prime}$ unitary matrix $\timeevop\left(t\from0\right)$
that can serve as the system's time-evolution operator. This argument
gives a novel and compelling first-principles explanation of why unitary
time evolution plays such a special role in quantum theory.

Crucially, for $N>2$, an $N\times N$ unistochastic matrix will not
generally be orthostochastic. An orthostochastic matrix is a unistochastic
matrix for which the underlying unitary matrix can be taken to have
only real-valued entries and therefore to form an orthogonal matrix.
Hence, to exploit the stochastic-to-quantum side of the correspondence,
with unitary time evolution, it will be necessary to introduce the
complex numbers, or an algebraic construct isomorphic to them. Of
course, from the standpoint of the stochastic\textendash quantum correspondence,
Hilbert spaces are mere mathematical fictions anyway, so introducing
the complex numbers to define the Hilbert-space picture is a harmless
move. The complex numbers also provide important benefits and resources,
like the spectral theorem, symmetry generators, Hamiltonians, energy
eigenvalues, stationary states, the Schrödinger equation, the uncertainty
principle, spinors, and more.

As it turns out, to implement time-reversal transformations, one again
needs to introduce the complex-conjugation operator $K$, as defined
in \eqref{eq:DefComplexConjOperator}, which combines with $i$ and
$iK$ to form the pseudo-quaternions, as in \eqref{eq:ElementaryPseudoQuaternionsAlgebra}.
In a sense, then, the Hilbert spaces of quantum systems are actually
defined over the pseudo-quaternions, rather than just over the complex
numbers, although observables are typically constructed out of only
the complex numbers.

After encoding the initial probabilities $p_{i}\left(0\right)$ as
the diagonal entries of an initial density matrix $\densitymatrix\left(0\right)$
that is otherwise filled with $0$s, one can go on to define a time-evolving
density matrix $\densitymatrix\left(t\right)\defeq\timeevop\left(t\from0\right)\densitymatrix\left(0\right)\timeevop^{\adj}\left(t\from0\right)$,
as well as the rest of the Hilbert-space picture, as detailed in other
work (Barandes 2025a)\nocite{Barandes:2025tsqc}. If the density
matrix is rank-one, then there exists an $N\times1$ state vector
or wave function $\ket{\Psi\left(t\right)}=\timeevop\left(t\from0\right)\ket{\Psi\left(0\right)}$
that gives a simple outer-product factorization 
\begin{equation}
\densitymatrix\left(t\right)=\ket{\Psi\left(t\right)}\bra{\Psi\left(t\right)},\label{eq:RankOneDensityMatrixFromStateVector}
\end{equation}
 where $\bra{\Psi\left(t\right)}\defeq\ket{\Psi\left(t\right)}^{\adj}$.
Assuming that the unitary time evolution is sufficiently smooth as
a function of $t$, the wave function then satisfies the Schrödinger
equation \eqref{eq:SchroEq}, 
\[
i\frac{\partial\ket{\Psi\left(t\right)}}{\partial t}=H\left(t\right)\ket{\Psi\left(t\right)},
\]
 for a self-adjoint Hamiltonian $H\left(t\right)=H^{\adj}\left(t\right)$
defined in keeping with Stone's theorem (Stone 1930)\nocite{Stone:1930ltihs}
in terms of the time-evolution operator $\timeevop\left(t\from0\right)$
according to 
\begin{equation}
H\left(t\right)\defeq i\frac{\partial\timeevop\left(t\from0\right)}{\partial t}\timeevop^{\adj}\left(t\from0\right).\label{eq:DefQuantumHamiltonian}
\end{equation}

The unitarily-time-evolving Hilbert-space formalism therefore yields
a \textquoteleft divisible\textquoteright{} Markovian embedding of
the original indivisible stochastic process, with a first-order differential
equation for the time evolution. Moreover, the linearity of the Schrödinger
equation now has a clear explanation: As one can check, it ultimately
descends from the linearity of the law of total probability \eqref{eq:LawOfTotalProbabilityFromTransitionMatrix}
connecting standalone probability distributions via the transition
matrix.

From the perspective of this formulation, one sees that wave functions
and the Schrödinger equation are secondary pieces of derived mathematics
and not the primary ontological furniture of quantum systems. In particular,
wave functions, like the luminiferous aether or Magritte's famous
pipe (Magritte 1929)\nocite{Magritte:1929ltdittoi} are not physical
or ontic. Wave functions are not even entirely epistemic, but encode
a blend of epistemic and nomological information, so they are not
well-classified according to the Harrigan-Spekkens \textquoteleft ontological
models\textquoteright{} framework (Harrigan, Spekkens 2010)\nocite{HarriganSpekkens:2010eievqs}.\footnote{Indeed, Harrigan and Spekkens expressly acknowledge that their ontic\textendash epistemic
dichotomy \textquotedblleft is presumably a category mistake\textquotedblright{}
in such cases.}

To establish the quantum-to-stochastic direction of the stochastic\textendash quantum
correspondence, one begins with a unitarily evolving quantum system,
with an $N$-dimensional Hilbert space $\hilbspace$. After picking
a convenient orthonormal basis (which is \emph{also} a step used to
define the path-integral formulation), one proceeds to define an indivisible
stochastic process via introducing a manifestly unistochastic transition
matrix $\stochasticmatrix\left(t\from0\right)$ according to the rule
\begin{equation}
\stochasticmatrix_{ij}\defeq\verts{\timeevop_{ij}\left(t\from0\right)}^{2}.\label{eq:UnistochasticTransitionMatrixFromModSqTimeEvOp}
\end{equation}

One might initially be concerned that this formula erases phase information
from $\timeevop\left(t\from0\right)$, as in \eqref{eq:ModGeneralComplexNumberPolar}.
However, this phase information does not matter as long as one remembers
that all empirical results come from measurement processes and one
is careful to include measuring devices and environments in models
of quantum systems. If, say, a measuring device is properly regarded
as a subsystem of the overall indivisible stochastic process, and
the chosen orthonormal basis for the overall system's Hilbert space
properly captures the measuring device's pointer variables, then,
by construction, the dynamics of the overarching indivisible stochastic
process will deliver the measuring device into its correct read-out
configurations with the correct measurement-outcome probabilities,
in accord with the Born rule. Keeping track of phase information in
$\timeevop\left(t\from0\right)$ is only necessary if one wishes to
\emph{exclude} the measuring device from the system, for reasons of
mathematical simplicity and convenience, in keeping with the usual
Dirac\textendash von Neumann axioms. A detailed treatment of the measurement
process within this framework is covered in other work (Barandes 2025a)\nocite{Barandes:2025tsqc}.

One therefore arrives at a comprehensive stochastic\textendash quantum
correspondence, according to which the Hilbert-space formalism is
a kind of a Markovian embedding, serving as a form of \textquoteleft analytical
mechanics\textquoteright{} for highly non-Markovian stochastic systems,
giving rise to an effective first-order, \textquoteleft divisible\textquoteright{}
differential equation.  Like any form of analytical mechanics, the
Hilbert-space formalism provides a powerful set of mathematical tools
for specifying microphysical laws in a systematic manner, for studying
dynamical symmetries, and for calculating predictions.

\section{Conclusion\label{sec:Conclusion}}

To summarize the arguments in this paper, one can regard a Hilbert-space
quantum system as a Markovian embedding for an indivisible stochastic
process, which essentially represents an equivalence class of non-Markovian
stochastic processes of arbitrarily high order. From this perspective,
the complex numbers are only necessary as part of reformulating an
indivisible stochastic process in the Hilbert-space picture with unitary
time evolution.

The claim that every quantum system is really an indivisible stochastic
process in disguise leads to a new interpretation of quantum theory,
naturally called \textquoteleft indivisible quantum theory,\textquoteright{}
which is based on simpler axioms than textbook quantum theory, does
away with treating superposition as a literal fact of physical objects
being in multiple configurations at once, arguably does not suffer
from a measurement problem, demotes wave functions from having a directly
physical or ontological status, and deflates a lot of the exotic talk
about quantum phenomena more generally. Finally, as argued elsewhere
(Barandes 2024)\nocite{Barandes:2024npfaclfoqt}, one can introduce
a plausible microphysical theory of causal influences according to
which indivisible quantum theory can be regarded as a causally local
theory.

\bibliographystyle{1_home_jacob_Documents_Work_My_Papers_2023-Stoc___ses_and_Quantum_Theory_custom-abbrvalphaurl}
\bibliography{0_home_jacob_Documents_Work_My_Papers_Bibliography_Global-Bibliography}

\end{document}